\documentstyle[10pt]{article}
\input{psfig.tex}
\begin{document}

\begin{center}
X-RAY EMISSION FROM A RADIO HOTSPOT IN 3C 390.3:\\
Evidence for the Deflection of a Radio Jet by a Neighboring Galaxy
\bigskip

D. E. Harris, Center for Astrophysics, 60 Garden St., Cambridge, MA 02138
USA, harris@cfa.harvard.edu\\
K.M. Leighly, Columbia Astrophysics Laboratory, 550 West 120th St., New York,
NY 10027 USA, leighly@ulisse.phys.columbia.edu\\
J.P. Leahy, University of Manchester, Nuffield Radio, Astronomy
Laboratories, Jodrell Bank, Macclesfield, Cheshire, SK11 9DL, England,
jpl@jb.man.ac.uk
 
\end{center}

\begin{abstract}
 
By summing a large number of ROSAT High Resolution Imager (HRI)
observations of the variable radio galaxy 3C 390.3, we demonstrate
that the X-ray emission associated with the northern radio lobe
(reported by Prieto 1997) can be identified with hotspot B.
None of the other hotspots have been detected.  We present evidence
that the anomalous X-ray emission is the consequence of a strong shock
produced where the northern radio jet impinges onto an external galaxy.
\end{abstract}

Subject headings: galaxies: active --- galaxies: individual (3C 390.3)
--- galaxies: jets --- radiation mechanisms: non-thermal --- radio
continuum: galaxies

\section{Introduction}
 
Very few radio hotspots in extragalactic sources have been detected in
the optical, and fewer still have been seen in the X-ray band.  The
problems posed by X-ray detection of radio features such as jets and
hotspots is of particular interest because the various emission models
used to explain them provide constraints on the physical parameters of
the sources beyond those normally obtainable from radio data alone.
The synchrotron model for knot A in the M87 jet requires electrons
with Lorentz energy factors, $\gamma=10^{7}$ and, for the
equipartition magnetic field strength, a half-life $\approx$~12
years (Biretta, Stern, \& Harris 1991).  The synchrotron
self-Compton model for the hotspots of Cygnus A (Harris, Carilli, \&
Perley 1994) yielded an average magnetic field strength in agreement
with the equipartition field (150 to 250 $\mu$G), whereas the
'Proton Induced Cascade' model (Mannheim, Krulls, \& Biermann 1991)
would require a much stronger B field, well in excess of 500
$\mu$G.
 
Prieto (1997) reported the detection of an X-ray feature associated
with the northern hotspots of 3C 390.3 with the ROSAT PSPC (11.5 ksec
exposure, producing $\sim$~80 net counts).  Leahy and Perley (1995)
presented VLA radio maps at 6 and 18cm, and we adopt their
nomenclature for the radio features.  Optical emission from hotspot B
(HS B, hereafter) had been reported twenty years ago with a brightness
of 25 B magnitudes arcsec$^{-2}$ (Saslaw, Tyson, \& Crane 1978) and
Prieto and Kotilainen (1997) have obtained V, R, and I magnitudes.
Keel and Martini (1995) presented an optical spectrum of a brighter
object about 5 arcsec to the north, which they found to be consistent
with the spectrum of a K4 (galactic) star.  However, as Prieto (1997)
points out, this object is outside HS B, and the astrometry rules this
out as a candidate for optical emission from the hotspot; we show
below that the object may be a member of a group of galaxies around 3C
390.3.
 
Leighly et al. (1997) report on a monitoring campaign of 3C 390.3
which involved a short observation with the ROSAT HRI every 3 days for
9 months.  We have stacked these data to obtain an exposure time of 197
ksec, and find a 5 $\sigma$ detection for HS B.  In this paper we
give the details of our measurements, consider various emission
mechanisms, and discuss the relevance of the galaxy adjacent to HS B.
What remains to be explained in detail is why HS B is detected in X-rays,
but the other hotspots (with higher radio luminosity) are below our
current X-ray detection limits.

\section{The Observational data and astrometry}
 
Leighly et al. (1997) report on the observational details.  For their
work on x-ray monitoring, all data intervals with excessively large
background levels were rejected, but we accepted all data that were
passed by the background rejection algorithms in the standard
processing.  We found that the resulting Point Response Function (PRF)
as described by the highly variable x-ray core had a Full Width
Half-Maximum (FWHM) size of $\sim$~10$^{\prime\prime}$ rather than the
value of $\sim$~5.5$^{\prime\prime}$ obtainable for data with good
aspect solutions.  We tried shifting each observation interval to
align the peak intensities but this did not appear to make a
significant improvement.
 
We then made a smoothed map using a Gaussian filter, precessed the
VLA\footnotemark~radio map to J2000, and shifted the X-ray map (less
than 1 arcsec) so as to align the radio and X-ray peaks of the core.
The results are shown in Figure 1.  It can be seen that the optical
position for the 19th magnitude object (which Keel and Martini
described as a K4 star) is well outside the hotspot, whereas the X-ray
emission is coincident with the radio emission.
 
\footnotetext{The National Radio Astronomy Observatory is a facility of
the National Science Foundation operated under cooperative agreement
by Associated Universities, Inc.}

\begin{figure}
\centerline{
\psfig{figure=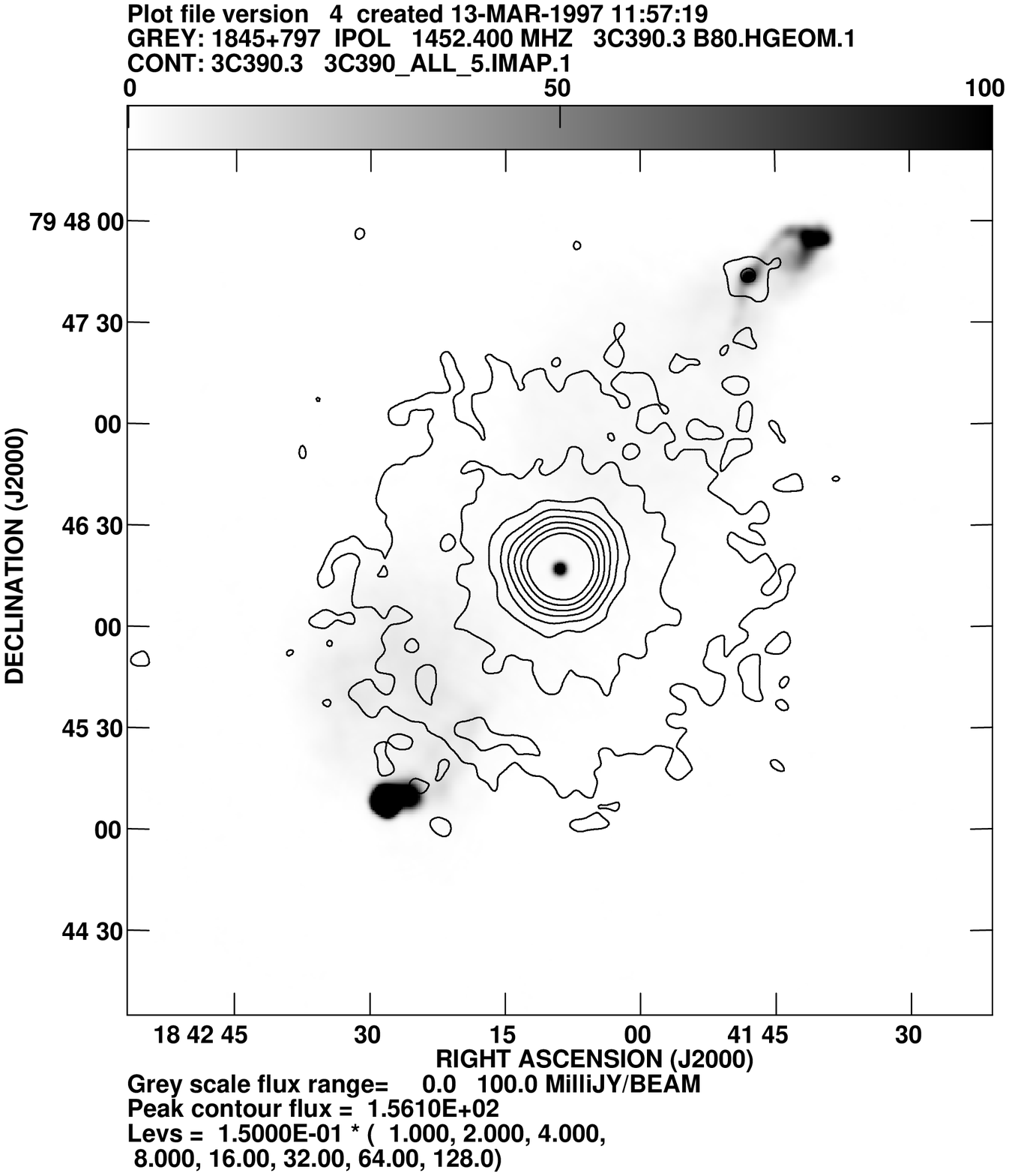,width=8cm}
\psfig{figure=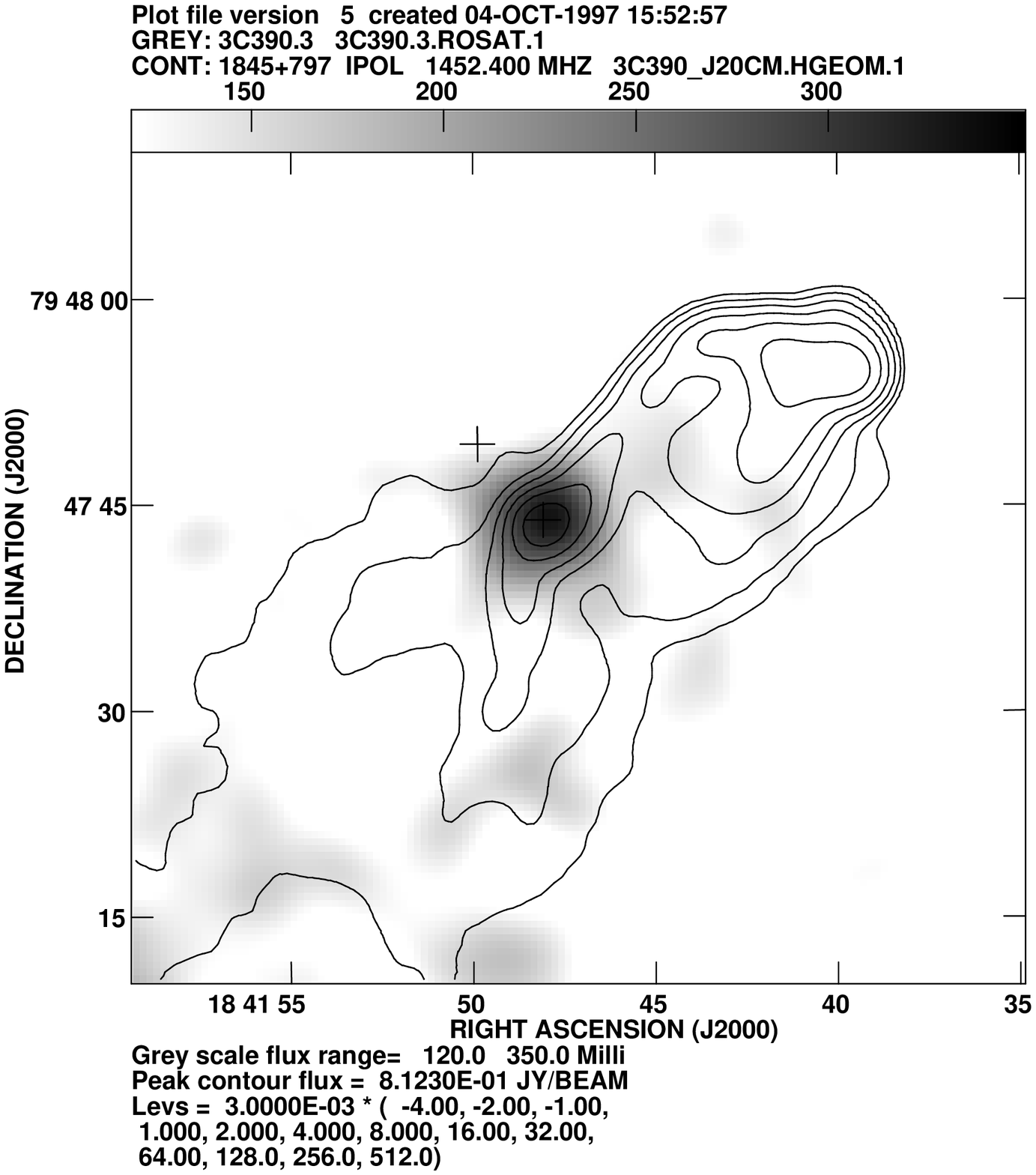,width=8cm}
}
\caption{\label{fig1}Radio and X-ray maps of 3C 390.3.  
(a) the entire source with the radio represented by the grey scale
(1452 MHz VLA map with a beam of 2.8 arcsec) and the X-ray shown as a
contour diagram.  The X-ray map was produced with a Gaussian smoothing
function of FWHM=5$^{\prime\prime}$.  The contours are logarithmic,
each increasing by a factor of two.  The first contour is at a
brightness level of 0.15 counts per pixel and the pixel size is 0.5
arcsec.
(b) the region around HS B.  The grey scale is now the 5$^{\prime\prime}$
smoothed X-ray map and the contours show the radio brightness.  The
contours are again logarithmic (factors of two) with the first level at
3 mJy/beam.  The cross shows the approximate position of the 19.6
magnitude object from Saslaw et al. (1978).}
\end{figure}

\section{Intensity Measurements}

Because of the large 'beam size' of the ROSAT HRI, we can measure only
three distinct hotspots: North preceding (Np) encompasses A, F, and N4
of Leahy and Perley (1995), and South following (Sf) which consists of
D, E, G, and the ``tail" (ibid.).  Note that radio features ``N1" and
``N2" are located within the measuring aperture of the HRI for HS B,
but there is no evidence that they contribute to the observed X-ray
emission (Figure 1b).
 
The X-ray intensities of hotspots Np and Sf were measured with a
circular aperture of radius 10$^{\prime\prime}$ and a background
annulus from 11$^{\prime\prime}$ to 17$^{\prime\prime}$.  Both values
were slightly negative; their entries in Table 1 are 2 $\sigma$ upper
limits.

The X-ray intensity of HS B was estimated by measuring the net count
rate in a circle of radius 10$^{\prime\prime}$, with a 10\% scattering
correction.  Background estimates were obtained from annuli centered
on the hotspot position with inner and outer radii of
11$^{\prime\prime}$ to 17$^{\prime\prime}$, 12$^{\prime\prime}$ to
20$^{\prime\prime}$, and 20$^{\prime\prime}$ to 30$^{\prime\prime}$.
The final countrate is 5$\times10^{-4}$~counts s$^{-1}$ (with a 1
$\sigma$ uncertainty of 20\%).
 
The countrate was converted to flux via the task 'hxflux' in
IRAF/PROS.  We assumed a power law spectrum (flux density, S $\propto
\nu^{-\alpha}$) with $\alpha$ = 0.8 and the Galactic value of log NH =
20.62.  The final value is 2.9$\times10^{-14}$ erg cm$^{-2}$ s$^{-1}$
for the 0.1 to 2.4 keV ROSAT bandpass.  Note that this value is
significantly lower than that published by Prieto (1997): S(1keV) = 13
$\pm$ 4 nJy compared to our value of 4.2 $\pm$ 0.8 nJy.  The actual
measuring aperture used by Prieto for the 80 net counts found for the
'northern extension' on the PSPC map was not given so we cannot make a
corresponding measurement on the HRI map.

\bigskip

\begin{center}
\begin{tabular}{lcclllc}
\multicolumn{7}{c}{Table 1. Optical and X-ray Flux Densities of the
Hotspots} \\
&&&&& \\ 
Feature	&RA&DEC	&Apert.&log $\nu$	&\phantom{$<$ }log S	&+/- \\
&\multicolumn{2}{c}{(J2000)}	&Size&(Hz)		&\phantom{$<$ }(cgs)	& \\
 &&&&&& \\
HS Np&18 41 39.84&79 47 54.7&r=1$^{\prime\prime}$&14.5823	      &$<$ -28.848& \\
&&&r=10$^{\prime\prime}$&17.383	      &$<$ -31.837& \\
&&&&&& \\
HS B&18 41 48.05 &79 47 43.6&r=1$^{\prime\prime}$&14.5823	&\phantom{$<$
}-28.488 &0.16 \\
&&&r=1$^{\prime\prime}$&14.6536	&\phantom{$<$ }-28.667 &0.12 \\
&&&r=1$^{\prime\prime}$&14.740	&\phantom{$<$ }-28.736 &0.16 \\
&&&...&14.834	&\phantom{$<$ }-29.097  &0.3 \\
&&&r=10$^{\prime\prime}$&17.383	&\phantom{$<$ }-31.377 &0.09 \\
&&&&&& \\
HS Sf&18 42 28.54&79 45 06.9&r=10$^{\prime\prime}$&17.383	&$<$ -31.837&
\end{tabular}
\end{center}

\begin{quote}
Notes to Table 1:

The cgs units for the flux density column are erg~cm$^{-2}$ s$^{-1}$
Hz$^{-1}$.  The error on our X-ray measurement is 1 $\sigma$.  The
optical B band is taken to be centered at $\nu = 6.8\times10^{14}$ Hz.
1 keV corresponds to 2.418 $\times 10^{17}$ Hz.  The optical flux
densities with r=1$^{\prime\prime}$ apertures are based on the
magnitudes published by Prieto and Kotilainen (1997).  Note that since
we do not know the effective bandwidths of their filters, the
conversion to flux density (based on standard filters) may not be
precise.
\end{quote}

For the optical intensity, we estimated a flux density from the
published remarks of Saslaw et al. (1978).  Since they only quote a
value of 25 B magnitudes per square arcsec, we have taken the
uncertainty to be that found by using integration areas of 1 to 3
square arcsecs.  Conversion to flux density followed the standard
prescriptions and no reddening corrections were used.  The magnitudes
given by Prieto and Kotilainen (1997) were also converted to flux
densities with the usual factors (see notes to Table 1).

\bigskip

\begin{center}
\begin{tabular}{cclccc}
\multicolumn{6}{c}{Table 2. Peak Radio Flux Densities for the
Hotspots} \\
&&&&& \\ 
 
Freq.	  &Beamsize	&VLA	&HS A	&HS B	&Sf	\\
	  &(arcsec)	&config.	&(mJy/b)	&(mJy/b)	&(mJy/b)	\\
&&&&& \\
1.5 GHz	  &2.8		&A+B+C	 &166.6	&163.4	 &815.1	\\
5.0 GHz	  &2.8		&B+C+D	  &62.4	 &66.6	 &346.9	\\
8.3 GHz	  &2.8		&C+D	  &44.4	 &49.2	 &246.4	\\
 &&&&& \\
4.8 GHz	  &0.33		&A	   &2.3	  &7.5	  &33.3
\end{tabular}
\end{center}

\begin{quote}
Notes to Table 2:

Uncertainties in the flux densities are generally dominated by the
overall flux calibration and are estimated to be $\pm$5\%.

'Sf' denotes 'South following' and includes hotspots D, E, and G.  At
0.33$^{\prime\prime}$ resolution, these hotspots are separable and
their peak flux densities are 21.0, 33.3, and 7.1 mJy, respectively.
Also at this resolution, hotspot F is distinct from HS A, and has a
peak value of 1.1 mJy.
\end{quote}

Comparative radio flux densities are more difficult to measure because
the high brightness features of hotspot B are surrounded by lower
brightness areas which probably do not contribute to the observed
optical and X-ray emissions.  In Table 2 we give peak values of flux
density measured from VLA maps.  The entries with the
2.8$^{\prime\prime}$ beam size are a reasonable approximation to the
total flux density of the hotspot because the brighter parts are
essentially unresolved at this resolution.

The optical, X-ray, and 2.8$^{\prime\prime}$ resolution radio flux
densities are plotted in figure 2.  The radio spectral indices of A,
B, and SF are 0.77, 0.70, and 0.70, respectively.  The radio/X-ray
value for HS B is 0.95.
 
\begin{figure}
\centerline{
\psfig{figure=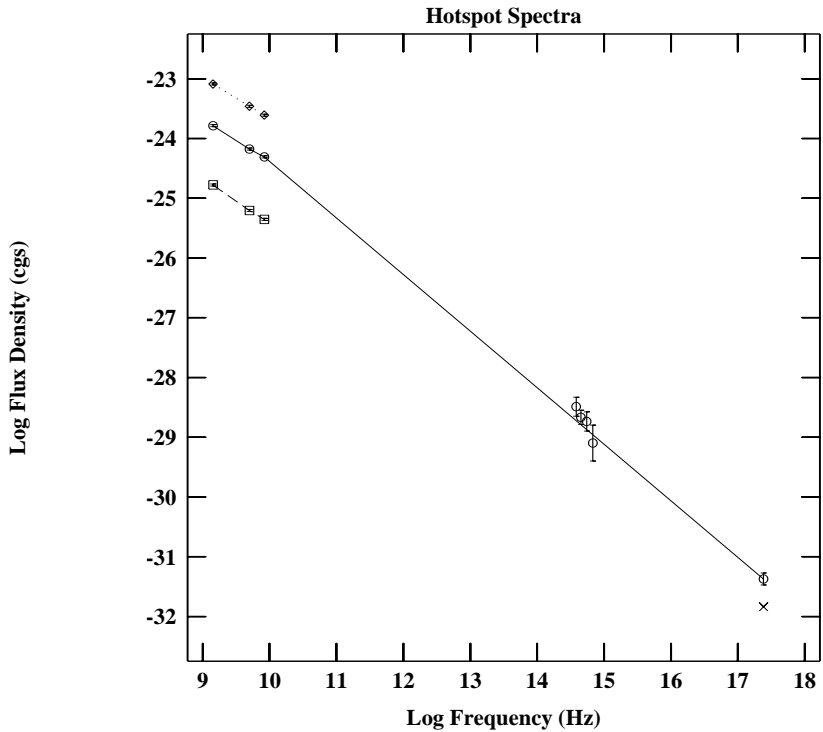,width=15cm,angle=-90}
} 
\caption{\label{fig2}Hotspot spectra from radio to X-rays.  The radio
points come from Table 2 (the 2.8$^{\prime\prime}$ beamsize).  The
optical and X-ray data are from Table 1.  The only X-ray detection is
for HS B; the cross just below that point is an upper limit for both
Np and Sf.  Circles are for HS B with a solid line connecting the
radio and X-ray points.  The dashed line between the squares are for
hotspot A (the Np hotspot) which, to avoid confusion with HS B, are
plotted a factor of 10 below the actual values. The dotted line and
diamonds are for the South following hotspot.}
\end{figure}

\section{The X-ray Emission Mechanisms for Hotspot B}

There are two obvious questions relating to the emission process of HS
B: ``Why is B detected whereas the (radio) brighter hotspots are
not?'' and ``Does the external galaxy just to the north of HS B cause
the radio jet to deflect, thereby engendering a shock capable of
producing the extremely energetic electrons required to produce X-ray
synchrotron emission in magnetic fields of order 50 $\mu$G?''
 
Prieto and Kotilainen (1997) give an upper limit of $m_{I} > 23$ for
hotspot A and Saslaw et al. (1978) state that no optical emission is
detected from the Np hotspot, but that the Sf hotspot location is
somewhat contaminated by a nearby star.  The radio spectrum of HS B is
essentially identical to that of Sf; it is only in the optical and
X-ray bands that HS B becomes exceptional.
 
We have also examined the radio surface brightnesses to see if there
is any significant difference between HS B and the others.  Although
HS B is brighter than HS A at the best radio resolution (0.33$^{\prime\prime}$), the
southern hotspots are brighter than HS B.

\subsection{Thermal Model}

For an order of magnitude estimate of the physical parameters required
to produce thermal bremsstrahlung emission, we take a uniform density
gas with a volume equal to that of a sphere of radius
3$^{\prime\prime}$ (characteristic of the ROSAT HRI point spread
function).  We find a required density of 0.07 cm$^{-3}$; a total mass
of 7~$\times~10^{8}$ M$_{\odot}$; a cooling time of 120 million years; and
a pressure of 9 $\times~10^{-10}$ dyne~ cm$^{-2}$.  These values were
derived for kT=4keV (4.4$\times~10^{7}$K); for gas temperatures
significantly different than 4 keV, the density and total mass would
have to be larger to produce the observed intensity.  The pressure and
cooling time increase for temperatures greater than 4 keV, but
decrease for cooler temperatures.

We believe our astrometry is accurate to better than
1$^{\prime\prime}$ and thus it appears highly probable that the X-ray
emission arises from the same volume as the radio (and optical)
emission.  While one might expect a high density (thermal) gas to be
associated with a shock front, there should then have been a clear
offset between the radio and X-ray positions.  Enhanced thermal
emission would be expected to the Northwest of the sharp edge of the
radio hotspot.  Since high brightness radio features generally exclude
ambient thermal material, the only geometry for a thermal origin of
the X-rays would be that of a thin sheath surrounding the hotspot.  We
suspect that this situation is also untenable because of the absence
of any Faraday effects.  Figure 6 of Leahy and Perley (1995) shows
that the rotation measure of hotspot B attains only very small values,
consistent with those of the other hotspots.  See also Harris et
al. (1994) for a similar discussion on the hotspots of Cygnus A.

\subsection{Synchrotron model}
 
The synchrotron model is based on the coincidence of the optical and
X-ray emitting volumes with that of the radio (fig 1b).  Although it
is obvious that a break in the spectrum is required, it is not known
if this occurs at a relatively discrete frequency or if there is a
more gradual curvature to the spectrum.  For the purposes of
estimating the usual synchrotron parameters, we assume that the
spectrum can be described by a power law with $\alpha$ = 0.70 between
0.1 and 10 GHz, and a second power law with $\alpha$ = 0.95 between 10
GHz and $10^{18}$ Hz (4.1 keV).  To obtain the magnetic field strength
for minimum energy, we assume the emitting volume to be a cylinder
with radius 0.5$^{\prime\prime}$ and length 1.3$^{\prime\prime}$
(estimated from the 5 GHz maps of Leahy \& Perley 1995).  The results
are given in Table 3.  None of these parameters are exceptional for
radio hotspots; the short lifetime of the highest energy electrons is
typical for any synchrotron model of X-ray emission.

\bigskip

\begin{center}
\begin{tabular}{lcl}
\multicolumn{3}{c}{Table 3. Synchrotron Parameters} \\
&& \\
Magnetic field for minimum pressure, B(min)&\hspace*{.5in}& 44
$\mu$G \\
Lorentz energy factor of electron 
	emitting X-rays, $\gamma$(max)&&7.5$\times 10^{7}$ \\
Halflife for $\gamma$(max) electrons in 44 $\mu$G field&&57 yr \\ 
Non thermal pressure, P(min)&&2$\times 10^{-10}$dyne cm$^{-2}$
\end{tabular}
\end{center}

\subsection{Synchrotron Self-Compton model}
 
From the observed radio volume and spectrum, we calculate the photon
energy density to be of order 10$^{-12}$ erg cm$^{-3}$ (the cosmic
microwave background provides $\sim$~4$\times 10^{-13}$).  For this
estimate, we assume that the photon spectrum observed in the radio
extrapolates up to $10^{13}$ Hz, thereafter dropping rapidly through
the optical points.  This together with the required electron spectrum
for B(min) provides an estimate for Synchrotron Self-Compton (SSC)
emission which is 3 orders of magnitude below that observed.  Even if
we have over estimated the volume by a considerable factor, it appears
that the SSC model is untenable.

\section{Discussion}

There are two aspects of HS B which suggest that the existence of the
hotspot at this location is caused by the external galaxy.  We have
obtained a 15 minute spectrum from the MMT (Blue channel; 300 line/mm,
with a slit width of 1.5$^{\prime\prime}$).  Although the s/n is too
low to allow a detailed analysis, the 4000 angstrom break is
discernible and using the galaxy template method, a redshift of 0.0527
+/- 0.0003 is found.  This value is about 840 km/s less than that of
3C 390.3 (z=0.0555, Eracleous \& Halpern 1998) and demonstrates that
the galaxy could be a member of the 3C 390.3 group.  Leahy and Perley
(1995) remarked that if this object were to be in the 3C 390.3 system,
it would be a 0.03L* dwarf. Other probable members of the group are
visible on the CCD image published by Baum et al. (1988).  On that
same CCD (Baum's figure 51), this galaxy is in the upper right corner
and appears to be resolved with a major axis in PA $\sim$~-35 degrees
whereas the foreground stars all have circular contours.  The deeper
images published by Prieto and Kotilainen (1997) leave no doubt that
the object is a galaxy and not a foreground star.
 
The second noteworthy feature of HS B is that the largest gradient in
radio surface brightness occurs on the 'outside' of the hotspot,
i.e. that bordering the dwarf galaxy.  This is similar to generic
hotspot behavior and is in sharp contrast to the X-ray emitting knot A
in the radio jet of M87 where the steep gradient occurs at the
upstream edge, facing the nucleus.  The best resolution radio map
(0.33 arcsec beam at 6cm) currently available is that published in
figure 12 of Leahy and Perley (1995) where it can also be seen that
the magnetic field is well aligned along the outside edge of the
hotspot.
 
If this scenario can be sustained, we suggest that the jet meets a
higher density region (the ISM of the dwarf galaxy?) and is deflected
by $\sim$~40 deg (in projection).  The resulting shock produces the
characteristic randomization of the relativistic electrons leading to
the radio hotspot.  Similar behavior has recently been reported for 3C
34 (Best, Longair, \& R\"{o}ttgering 1997) and 3C 371 ( Nilsson et
al. 1997).

\section{acknowledgments}
 
We thank J. Halpern for helpful discussions and Jane Dennett-Thorpe
for the X- and C-band low-resolution images.  H. Glasser, a
Haverford College extern, participated in the initial stages of the
X-ray reductions and evaluation.  The optical observations reported in
this paper were obtained at the Multiple Mirror Telescope Observatory,
a joint facility of the University of Arizona and the Smithsonian
Institution.  It is a pleasure to thank J. Huchra, S. Tocarz, and
J. Mader for assistance in reducing the MMT spectrum and P. Berlind
for making the observation.  This work was partially supported by
NASA contract NAS5-30934.
 
\section{references}
 
Baum, S. A., Heckman, T., Bridle, A., van Breugel, W., \& Miley, G. 1988, ApJS 68, 643\\
Best, P. N., Longair, M. S., \& R\"{o}ttgering, H. J. A. 1997, MNRAS 286, 785\\
Biretta, J.A., Stern, C.P., and Harris, D.E. 1991, AJ 101, 1632\\
Eracleous, M., \& Halpern, J. P. 1998, in preparation\\
Harris, D. E., Carilli, C. L., \& Perley, R. A. 1994, Nature 367, 713\\
Keel, W. C. \& Martini, P. 1995, AJ 109, 2305\\
Leahy, J. P. \& Perley, R. A. 1995, MNRAS 277, 1097\\
Leighly, K. M. et al. 1997, ApJ, 483, 767\\
Mannheim, K., Krulls, W. M., \& Biermann, P. L. 1991, A\&A 251, 723\\
Nilsson, K., Heidt, J., Pursimo, T., Sillanp\"{a}\"{a}, A., Takalo, L. O. \& J\"{a}ger, K. 1997, ApJ 484, L107\\
Prieto, M. A. 1997, MNRAS 284, 627\\
Prieto, M. A. \& Kotilainen, J. K. 1997, ApJ 491, L77\\
Saslaw, W. C., Tyson, J. A., \& Crane, P. 1978 ApJ 222, 435\\

\end{document}